\documentclass[12pt]{iopart}

\usepackage{iopams}
\usepackage{graphicx}
\usepackage{bm}
\newcommand{\ket}[1]{|#1\rangle}

\begin{document}

\title{Two-qubit entanglement dynamics for two different non-Markovian environments}

\author{Bruno Bellomo$^1$, Rosario Lo Franco$^1$, Sabrina Maniscalco$^2$ and Giuseppe Compagno$^1$}
\address{$^1$CNISM and Dipartimento di Scienze Fisiche ed Astronomiche,
Universit\`{a} di Palermo, via Archirafi 36, 90123 Palermo, Italy}
\address{$^2$Department of Physics and Astronomy, University of Turku, Turun yliopisto, FIN-20014 Turku, Finland}
\ead{lofranco@fisica.unipa.it}

\begin{abstract}
We study the time behavior of entanglement between two noninteracting qubits each immersed in its own environment for two different non-Markovian conditions: a high-$Q$ cavity slightly off-resonant with the qubit transition frequency and a nonperfect photonic band-gap, respectively. We find that revivals and retardation of entanglement loss may occur by adjusting the cavity-qubit detuning, in the first case, while partial entanglement trapping occurs in non-ideal photonic-band gap.
\end{abstract}

\maketitle

\section{\label{intro}Introduction}
Real quantum systems unavoidably interact with their surroundings undergoing a consequent decoherence and entanglement loss \cite{petru}. It is known that two entangled qubits embedded in Markovian (memoryless) environments may become completely disentangled at a finite time, in spite of an exponential decay of the single qubit coherence \cite{diosi,yu2004PRL}. This entanglement sudden death (ESD, or early-stage disentanglement) which has been experimentally revealed \cite{almeida2007Science,kimble2007PRL}, puts a serious limit to the storage times of entanglement for practical purposes, e.g., for the realization of quantum memory banks \cite{yu2009Science}. A realistic quantum computer will probably have to take into account this quantum dynamical drawback. It is therefore important to study the possibile physical conditions where entanglement can be maintained.

Entanglement losses during the evolution crucially depend on the particular noise acting on the system. Under Markovian-noise conditions the quantum process is typically irreversible. Differently, structured environments or strong coupling can give rise to non-Markovian noise (environment with memory) whose effects on the entanglement dynamics is currently subjected to investigation \cite{yu2009Science}. In this context, the cases of two noninteracting qubits embedded either in separated high-$Q$ cavities \cite{bellomo2007PRL,bellomo2008PRA} or in a common cavity \cite{mazzola2009PRA} supporting a mode resonant with the qubit transition frequency have been analyzed, finding that revivals of the initial two-qubit entanglement can occur. When the qubits share a common environment, it has been also shown that entanglement can be preserved by means of quantum Zeno effect \cite{maniscalco2008PRL}. Moreover, entanglement trapping is achievable when two independent qubits are embedded in an ideal photonic-band gap (PBG) material (photonic crystal) \cite{bellomo2008trapping,wang2008PRA}.

The aim of this paper is therefore to deepen the analysis of entanglement evolution in non-Markovian environments, considering in particular two different effective spectral conditions of the environment-qubit system simulating, respectively, a cavity with a mode slightly off-resonant with the qubit transition frequency and a nonperfect photonic band-gap at the qubit transition frequency. In this paper we highlight the differences with previous studies and we discuss the optimal physical parameters for observing entanglement revivals and for preservation of entanglement.

\section{\label{model}Model}
We consider a system composed by two parts $\tilde{S}=\tilde{A},\tilde{B}$, each one consisting of a two-level system (qubit) $S=A, B$ interacting with a reservoir $R_S=R_A, R_B$.

The single part $\tilde{S}$ ``qubit S +reservoir $R_S$'' is described by the Hamiltonian
\begin{equation}\label{Hamiltonian}
\hat{H}_{\tilde{S}}=\hbar \omega_0\hat{\sigma}_+\hat{\sigma}_-+\sum_k\hbar\left[\omega_k \hat{b}_k^\dag \hat{b}_k+\left(g_k \hat{\sigma}_+\hat{b}_k  + g_k^* \hat{\sigma}_-\hat{b}_k^\dag\right)\right],
\end{equation}
where $\omega_0$ is the transition frequency and $\sigma_ \pm$ are the qubit raising and lowering
operators, $b_k^\dag $, $b_k $ are the creation and annihilation operators and $g_k$ the coupling
constant of the mode $k$ with frequency $\omega_k$. When the environment is at zero temperature the single-qubit reduced density matrix $\hat{\rho}^S(t)$ can be written, in the basis $\{\ket{1},\ket{0}\}$, as \cite{petru}
\begin{equation}\label{roS}
\hat{\rho}^S(t)=\left(%
\begin{array}{cc}
\rho^S_{11}(0)|q(t)|^2  & \rho^S_{10}(0)q(t)\\\\
\rho^S_{01}(0)q^*(t)  & \rho^S_{00}(0)+ \rho^S_{11}(0)(1-|q(t)|^2) \\
\end{array}\right).
\end{equation}
From the equation above it is readily seen that the single-qubit dynamics depends only on the function $q(t)$ that in turn is determined by the reservoir spectral density. Indeed, $q(t)$ obeys the differential equation $\dot{q}(t)=-\int_0^t\mathrm{d}t_1 f(t-t_1)q(t_1)$, and the correlation function $f(t-t_1)$ is related to the spectral density $J(\omega)$ of the reservoir by $f(t-t_1)=\int \mathrm{d}\omega J(\omega)\exp[i(\omega_0-\omega)(t-t_1)]$. The solution of the associated algebraic equation for $\dot{q}(t)$, obtained trough Laplace transforms, is $\bar{q}(s)=q(0)/[s+\bar{f}(s)]$, where $\bar{q}(s)$ and $\bar{f}(s)$ are the Laplace transforms of $q(t)$ and $f(t-t_1)$.

A crucial quantity for our study is the explicit analytic expression of the two-qubit reduced density matrix at the time $t$. For the system of noninteracting qubits in separated environments here considered this can be obtained by a procedure based on the knowledge of the single-qubit dynamics \cite{bellomo2007PRL}. In fact, given the time-dependent single-qubit density matrix elements as $\rho^{A}_{ii'}(t)=\sum_{ll'}A_{ii'}^{ll'}(t)\rho^{A}_{ll''}(0)$, $\rho^{B}_{jj'}(t)=\sum_{mm'}B_{jj'}^{mm'}(t)\rho^{B}_{mm'}(0)$, the time-dependent two-qubit density matrix elements are \cite{bellomo2007PRL}
\begin{equation}\label{totalevo}
\rho_{ii',jj'}(t)=\sum_{ll',mm'}A_{ii'}^{ll'}(t)B_{jj'}^{mm'}(t)\rho^{}_{ll',mm'}(0),
\end{equation}
where $i,j,l,m=0,1$. The two-qubit density matrix $\hat{\rho}(t)$ is thus obtained by means of Eq.~(\ref{totalevo}) for an arbitrary two-qubit initial condition, its elements depending only on their initial values and on the function $q(t)$ \cite{bellomo2007PRL,bellomo2008PRA}. In the following these density matrix elements will be meant in the standard computational basis $\mathcal{B}=\{\ket{1}\equiv\ket{11},\ket{2}\equiv\ket{10}, \ket{3}\equiv\ket{01}, \ket{4}\equiv\ket{00}\}$.

\subsection{Initial states}
With regards to the initial state, we will limit our analysis to the case of initial pure Bell-like states
\begin{eqnarray}\label{Bell-likestates}
\ket{\Phi}=\alpha\ket{01}+\beta e^{i\delta}\ket{10},\quad\ket{\Psi}=\alpha\ket{00}+\beta e^{i\delta}\ket{11},
\end{eqnarray}
with $\alpha,\beta$ real and $\alpha^2+\beta^2=1$. For $\alpha=\pm\beta=1/\sqrt{2}$ these states coincide with the Bell states. Bell-like states have the property that their resulting density matrix has an X structure (only diagonal and antidiagonal density matrix elements different from zero). Under our dynamical conditions, the X structure is maintained during the two-qubit evolution, so that the two-qubit density matrix at time $t$ will also have a X structure.

\subsection{Concurrence}
In order to quantify the entanglement during the evolution of the bipartite system, we use the concurrence $C$ \cite{wootters1998PRL}. The concurrence at the time $t$ for an initial general X state can be easily computed by exploiting the fact that the X structure is preserved here and by using Eqs.~(\ref{roS}) and (\ref{totalevo}). The expression of the concurrence is given by \cite{yu5}
{\setlength\arraycolsep{1pt}\begin{eqnarray}\label{concurrence x state}
&&C_\rho^X(t)= 2\mathrm{max}\{0,K_1(t),K_2(t)\},\nonumber \\
&&K_1(t)=|q(t)|^2\Big\{\rho_{23}(0)-\sqrt{\rho_{11}(0)}[\rho_{44}(0)+\rho_{11}(0) \nonumber \\
&&\times(1-|q(t)|^2)^2 +(\rho_{22}(0)+\rho_{33}(0))(1-|q(t)|^2)]^\frac{1}{2}\Big\},\nonumber\\
&&K_2(t)=|q(t)|^2\Big[\rho_{14}(0)-\sqrt{\rho_{22}(0)+\rho_{11}(0)(1-|q(t)|^2)}\nonumber \\
&&\times\sqrt{\rho_{33}(0)+\rho_{11}(0)(1-|q(t)|^2)}\Big].
\end{eqnarray}}
These formulas are quite general since their form does not explicitly depend on the particular choice of the environment, but only on the Hamiltonian model of Eq.~(\ref{Hamiltonian}) and on the chosen initial state. The explicit time dependence of concurrence depends on the explicit form of the function $q(t)$ and thus it contains the information about the environment structure. In the following we shall consider two different environment structures with given spectral densities which in turn shall determine the explicit form of $q(t)$.

\section{\label{Different spectra} Spectral density effect on the entanglement dynamics}

We shall now analyze the evolution of concurrence for two different spectral densities: a single Lorentzian simulating a cavity with a mode nonresonant with the qubit transition frequency and a nonperfect photonic band-gap at the qubit transition frequency.

\subsection{\label{Single Lorentzian} Off-resonant high-$Q$ cavity}
As a first example we take the spectral distribution $J(\omega)$ of the electromagnetic field inside a high-$Q$ cavity supporting a mode detuned by $\Delta$ from the qubit (two-level atom) transition frequency $\omega_0$, resulting from the combination of the environment spectrum and the system-environment coupling. It has the Lorentzian form \cite{petru}
\begin{equation}\label{spectral density Single Lorentzian}
J(\omega)=\frac{1}{2\pi}\frac{\Gamma \lambda^2}{(\omega_0-\Delta-\omega)^2+\lambda^2},
\end{equation}
where $\Gamma$ is the qubit free space linewidth and $\lambda$ the spectral width of the coupling. The parameter $\lambda$ is then connected to the reservoir correlation time $\tau_B$ by the relation $\tau_B\approx \lambda^{-1}$. The relaxation timescale $\tau_R$ over which the state of the system changes is related to $\Gamma$ by $\tau_R\approx\Gamma^{-1}$. The correlation function corresponding to this $J(\omega)$ is $f(t-t_1)=\frac{\Gamma \lambda}{2}\exp[-(\lambda-i\Delta )(t-t_1)]$. Using this correlation function, the Laplace transform of $q(t)$ is $\bar{q}(s)=1/\left[s+\frac{1}{2}\frac{\Gamma \lambda}{s-(\lambda -i \Delta )}\right]$ and inverse Laplace transform finally gives
\begin{equation}\label{qnonresonantcavity}
q(t)=\mathrm{e}^{-\frac{\lambda-i \Delta}{2} t}\left[\cosh \left(\frac{d t}{2}\right)+\frac{\lambda -i\Delta }{d}
\sinh\left(\frac{dt}{2}\right)\right],
\end{equation}
where $d=\sqrt{(\lambda-i\Delta)^2-2\Gamma\lambda}$.

In the resonant limit, $\Delta=0$, the correlation function has an exponential form with $\lambda$ the decay rate. In the analysis of the function $q(t)$ of Eq.~(\ref{qnonresonantcavity}) with $\Delta=0$ a weak ($\Gamma<\lambda/2$) and a strong ($\Gamma>\lambda/2$) coupling regime can be distinguished  \cite{petru,maniscalco}. In the weak coupling regime the relaxation time is greater than the reservoir correlation time ($\tau_R>2\tau_B$) and the behavior of $q(t)$ is essentially a Markovian exponential decay controlled by $\Gamma$. In the strong coupling regime, the reservoir correlation time is greater or of the same order of the relaxation time ($\tau_R<2\tau_B$) and non-Markovian effects become relevant. Within this regime, $q(t)$ presents oscillations describing a quasi-coherent exchange of energy between the qubit and the reservoir.

\subsubsection{Entanglement dynamics.}
We now investigate the entanglement dynamics of the two-qubit system considering in detail the effect of the detuning $\Delta$, by using the concurrence obtained by Eq.~(\ref{concurrence x state}) with $q(t)$ given by Eq.~(\ref{qnonresonantcavity}).
\begin{figure}
\begin{center}
\includegraphics[width=8.1 cm, height=5.5 cm]{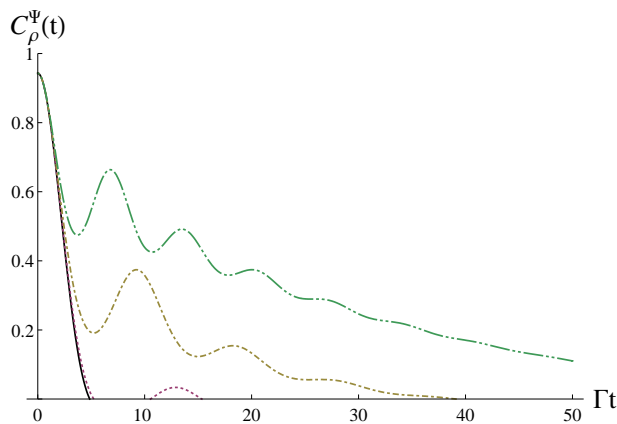}
\caption{\label{fig1}\footnotesize Nonresonant cavity: $\lambda=0.1\Gamma$. Concurrence as a function of the dimensionless quantities $\Gamma t$ starting from the initial EWL state $\hat{\rho}^\Psi(0)=|\Psi\rangle\langle  \Psi|$ with $\alpha=1/\sqrt{3}$ for different values of detuning $\Delta$: $\Delta=0$ (solid curve), $\Delta=2 \lambda$ (dotted curve), $\Delta=5 \lambda$ (long-short-dashed curve), $\Delta=8 \lambda$ (long-short-short-dashed curve).}
\end{center}
\end{figure}
The evolution of concurrence for various values of the detuning is shown in Fig.~\ref{fig1} when the cavity bandwidth $\lambda$ is smaller than the free-space atomic linewidth $\Gamma$ ($\lambda=0.1\Gamma$) and the initial state is not maximally entangled, in particular $\hat{\rho}^\Psi(0)=|\Psi\rangle \langle \Psi|$ with $\alpha=1/\sqrt{3}$. In the resonant case it is found, as known, that the state suffers ESD \cite{bellomo2007PRL}. However, from the plot one observes that increasing the detuning, the entanglement decay slows down. In particular, for $\Delta=2\lambda$ we have also revivals of entanglement after a finite period of time when the two qubits are not entangled. The phenomena of the slowing down of entanglement decay and of the entanglement revivals are a clear manifestation of the environment memory effects.

\subsection{\label{photonicbandgap}Nonperfect photonic band-gap}
As a second example of non-Markovian environment, we consider a spectral density of the form \cite{garraway1997PRA}
\begin{equation}\label{spectraldensitybandgap}
    J(\omega)=\frac{1}{2\pi}\left( \frac{\Gamma_1 \lambda_1^2}{(\omega-\omega_0)^2+\lambda_1^2}-\frac{\Gamma_2 \lambda_2^2}{(\omega-\omega_0)^2+\lambda_2^2}\right),
\end{equation}
which represents a Lorentzian with a dip used as a model to simulate the spontaneous decay of a qubit in a nonperfect photonic band gap (PBG). In Eq.~(\ref{spectraldensitybandgap}), $\lambda_1$ represents the bandwidth of the flat background continuum, $\lambda_2$ the width of the gap, and $\Gamma_1$ and $\Gamma_2$ the strength of the background and the gap, respectively. The spectral density must be positive, this implies $\Gamma_1\lambda_1^2>\Gamma_2\lambda_2^2$ (condition for $J(\omega)$ to be positive at large $\omega$) and $\Gamma_1>\Gamma_2$ (condition for $J(\omega)$ to be positive at the center of resonance). Combining these two relations the condition $\Gamma_1\lambda_1>\Gamma_2\lambda_2$ must be satisfied, this being also the condition for a localized dip \cite{garraway1997PRA}. In the case $\Gamma_1=\Gamma_2$ the spectral density reduces exactly to zero at the center of the gap ($\omega=\omega_0$), leading to population trapping \cite{garraway1997PRA}. For this form of $J(\omega)$, one obtains for the correlation function
$f(t-t_1)=\left( \Gamma_1\lambda_1 e^{-\lambda_1(t-t_1)}-\Gamma_2\lambda_2 e^{-\lambda_2(t-t_1)}\right)/2$ and the Laplace transform of $q(t)$ becomes
\begin{eqnarray}\label{equforp2}
\bar{q}(s) =\frac{(\lambda_1+s)(\lambda_2+s)}{s^3+s^2(\lambda_1+\lambda_2)+s(\lambda_1\lambda_2+\Lambda)+ \lambda_1\lambda_2\Gamma_\mathrm{d}},
\end{eqnarray}
where $\Lambda=(\Gamma_1\lambda_1-\Gamma_2\lambda_2)/2$ and $\Gamma_\mathrm{d}=(\Gamma_1-\Gamma_2)/2$. By inverting the Laplace transform one finally obtains
\begin{equation}\label{qbandgap}
q(t)=\sum_i\frac{u_i^2+u_i(\lambda_1+\lambda_2)+\lambda_1\lambda_2}{(u_i-u_j)(u_i-u_k)}\mathrm{e}^{u_it},
\end{equation}
where $i,j,k=1,2,3$ are all different indexes and $u_i$ are the three solutions of the third degree equation appearing in the denominator of Eq.~(\ref{equforp2}). The explicit expressions of $u_i$ are rather complex and we shall not report them here.

\subsubsection{Entanglement dynamics.}
We are now ready to analyze the entanglement dynamics of the two-qubit system in a nonperfect PBG as described by the spectral density of Eq.~(\ref{spectraldensitybandgap}), by using the concurrence of Eq.~(\ref{concurrence x state}) with $q(t)$ given by Eq.~(\ref{qbandgap}).
\begin{figure}
\begin{center}
\includegraphics[width=8.1 cm, height=5.5 cm]{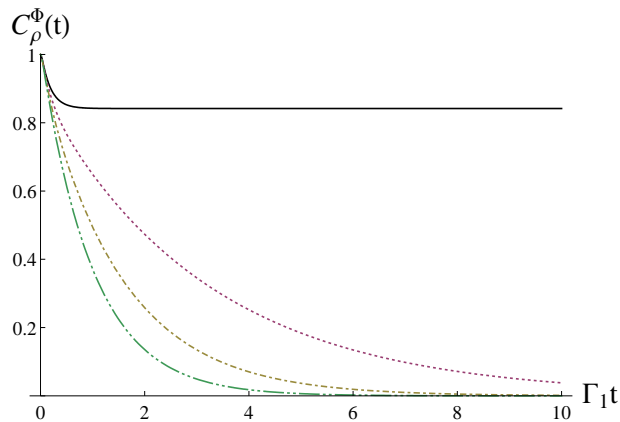}
\caption{\label{fig2}\footnotesize Nonperfect PBG case: $\lambda_1=10\lambda_2=50\Gamma_1$. Concurrence as a function of the dimensionless quantity $\Gamma_1 t$ starting from the initial state $\hat{\rho}^\Phi(0)=|\Phi\rangle \langle \Phi|$ with $\alpha=\beta=1/\sqrt{2}$ for different values of $\Gamma_2$:  $\Gamma_2=\Gamma_1$ (solid curve), $\Gamma_2=\Gamma_2/3$ (dotted curve), $\Gamma_2=2\Gamma_1/3$ (long-short-dashed curve), $\Gamma_2=0$ (long-short-short-dashed curve).}
\end{center}
\end{figure}
In Fig.~\ref{fig2} we investigate the evolution of concurrence for $\lambda_1=10\lambda_2=50\Gamma_1$ and varying $\Gamma_2$. As initial state we consider the Bell state $\hat{\rho}^\Phi(0)=|\Phi\rangle \langle \Phi|$ ($\alpha=\beta=1/\sqrt{2}$). For $\Gamma_2=\Gamma_1$, the spectral density goes to zero at the center of the gap and as a consequence we obtain entanglement trapping (similarly to what happens in Ref.~\cite{bellomo2008trapping}). The other curves are obtained for decreasing values of $\Gamma_2$: $\Gamma_2=2\Gamma_2/3$, $\Gamma_2=\Gamma_1/3$ and $\Gamma_2=0$. Smaller values of $\Gamma_2$ correspond to a smaller dip of the spectral density at $\omega=\omega_0$. In particular, for $\Gamma_2=0$, the shape of the spectral density is again a simple Lorentzian. In Fig.~\ref{fig2} the value of $\lambda_1$ is chosen so that we are in weak coupling regime (see Sec.\ref{Single Lorentzian}) and a Markovian decay occurs. Finally, the plot evidences how decreasing $\Gamma_2$ the trapping of entanglement is lost and the entanglement decay speeds up always more.

\section{Conclusions}
In this paper we have extended previous analysis on entanglement dynamics of two noninteracting qubits embedded in bosonic environments at zero-temperature. We have examined two different spectral densities corresponding to two different environments: the first case considered is  a Lorentzian spectrum representing a high-$Q$ cavity out of resonance with the qubit transition frequency; the second, a nonperfect photonic-band gap. The first case has allowed to analyze the role of the cavity-qubit detuning comparing it to the known resonant case. In particular, as expected, an increase of the entanglement lifetime is observed when the detuning is increased. On the other hand, the second spectral density has permitted the study of entanglement dynamics when ideal conditions of the photonic-band gap (spectral density exactly equal to zero for a frequency equal to the qubit transition frequency) are not satisfied. In this case, it has been found that increasing the value of the spectral density in the central frequency (qubit transition frequency) entanglement trapping, which is expected for the ideal case, vanishes while entanglement decay speeds up. This study has thus enlarged the knowledge of entanglement evolution under different non-Markovian conditions, providing more hints for future investigations on this topic.

R.L.F. (G.C.) acknowledges partial support by MIUR project II04C0E3F3 (II04C1AF4E) \textit{Collaborazioni Interuniversitarie ed Internazionali tipologia C}.

S. M. acknowledges financial support from the Turku Collegium of Science and Medicine, the Academy of Finland, the V\"{a}is\"{a}l\"{a} Foundation, the Magnus Ehrnrooth Foundation, and the Turku University Foundation.

\providecommand{\newblock}{}

\end{document}